\documentclass[%
 reprint, prl, amsmath,amssymb,superscriptaddress]{revtex4-1}

\usepackage{graphicx}
\usepackage{dcolumn}
\usepackage{bm}
\usepackage{SIunits}
\usepackage{braket}
\usepackage{amssymb}
\usepackage{color}

\newcommand{\nm}{\nano\metre}

\newcommand{\um}{\micro\metre}
\newcommand{\GHz}{\giga\hertz}
\newcommand{\MHz}{\mega\hertz}

\newcommand{\us}{\micro\second}
\newcommand{\ms}{\milli\second}
\newcommand{\mW}{\milli\watt}

\begin{document}


\title{Creation of ultracold $^{87}$Rb$^{133}$Cs molecules in the rovibrational
ground state}

\author{Peter K. Molony}
\author{Philip D. Gregory}
\author{Zhonghua Ji}
\author{Bo Lu}
\author{Michael P. K\"oppinger}
\affiliation{Joint Quantum Centre (JQC) Durham-Newcastle, Department of
Physics, Durham University, South Road, Durham DH1 3LE, United Kingdom}
\author{C.~Ruth~Le~Sueur}
\author{Caroline L. Blackley}
\author{Jeremy M. Hutson}
\affiliation{Joint Quantum Centre (JQC) Durham-Newcastle, Department of
Chemistry,
Durham University, South Road, Durham, DH1 3LE, United Kingdom}
\author{Simon L. Cornish}
\email{s.l.cornish@durham.ac.uk}
\affiliation{Joint Quantum Centre (JQC) Durham-Newcastle, Department of
Physics, Durham University, South Road, Durham DH1 3LE, United Kingdom}

\date{\today}


\begin{abstract}
We report the creation of a sample of over 1000 ultracold $^{87}$Rb$^{133}$Cs molecules
in the lowest rovibrational ground state, from an atomic mixture of $^{87}$Rb
and $^{133}$Cs, by magnetoassociation on an interspecies Feshbach resonance followed by
stimulated Raman adiabatic passage (STIRAP). We measure the binding energy of
the RbCs molecule to be $h
c~\times~\unit{3811.576(1)}{\reciprocal{\centi\metre}}$ and the $\ket{v''=0,
J''=0}$ to $\ket{v''=0, J''=2}$ splitting to be $h \times
\unit{2940.09(6)}{MHz}$. Stark spectroscopy of the rovibrational ground state
yields an electric dipole moment of $\unit{1.225(3)(8)}{D}$, where the values
in parentheses are the statistical and systematic uncertainties, respectively.
We can access a space-fixed dipole moment of 0.355(2)(4)~D, which is
substantially higher than in previous work.
\end{abstract}

\maketitle


The quest for ultracold samples of trapped polar molecules has attracted
considerable attention over the last decade \cite{L.D.Carr_NJP_2009,
D.S.Jin_CR_2012}. The permanent electric dipole moments of polar molecules give
rise to anisotropic, long-range dipole-dipole interactions which can be tuned
by applied electric fields \cite{M.Lemeshko_MP_2013}. This property, combined
with the exquisite control of ultracold systems, offers exciting prospects in
the fields of quantum controlled chemistry \cite{S.Ospelkaus_Science_2010,
R.V.Krems_PCCP_2008}, precision measurement \cite{V.V.Flambaum_PRL_2007,
T.A.Isaev_PRA_2010, J.J.Hudson_Nature_2011}, quantum computation
\cite{D.DeMille_PRL_2002} and quantum simulation \cite{L.Santos_PRL_2000,
M.A.Baranov_CR_2012}.

Direct cooling of molecules into the ultracold regime remains elusive, though
recent demonstrations of laser cooling show great
promise~\cite{E.S.Shuman_Nature_2010, M.T.Hummon_PRL_2013,
V.Zhelyazkova_PRA_2014}. An alternative approach is to form ultracold molecules
indirectly by association of pre-cooled
atoms~\cite{T.Kohler_RMP_2006,K.M.Jones_RMP_2006}. To date, the most successful
method has employed magnetoassociation on a Feshbach
resonance~\cite{T.Kohler_RMP_2006,C.Chin_RMP_2010} to produce weakly bound
molecules which are subsequently transferred to the rovibrational ground state
by stimulated Raman adiabatic passage (STIRAP)~\cite{K.Bergmann_PRA_1998}.
Although this technique has been successfully applied in homonuclear
Cs$_2$~\cite{J.G.Danzl_Nature_2010} and triplet
$^{87}$Rb$_2$~\cite{F.Lang_PRL_2008}, experiments exploring the role of
dipole-dipole interactions have been confined to the fermionic $^{40}$K$^{87}$Rb
molecule~\cite{K.-K.Ni_Science_2008}. However, KRb molecules are unstable, as
the exchange reaction $2\textrm{KRb}\rightarrow \textrm{K}_2+\textrm{Rb}_2$ is
exothermic~\cite{S.Ospelkaus_Science_2010}. This leads to significant loss of
the molecules~\cite{K.-K.Ni_Nature_2010}. Nevertheless, the confinement of the
molecules in a three-dimensional optical
lattice~\cite{M.H.G.Miranda_nature_2011} eliminates this reaction and has
allowed pioneering studies of dipolar spin-exchange
interactions~\cite{B.Yan_Nature_2013}.

Ground-state RbCs molecules offer an appealing alternative to KRb because both
the exchange and trimer formation reactions are endothermic~\cite{
P.S.Zuchowski_PRA_2010,Reactions}. The bosonic $^{87}$Rb$^{133}$Cs molecule
also offers a contrast with fermionic $^{40}$K$^{87}$Rb. Moreover, the large
predicted electric dipole moment of $\unit{1.28}{D}$~\cite{M.Aymar_JCP_2005} is
easily aligned in the laboratory frame, meaning that only modest electric
fields are required to realize significant dipole-dipole interactions.
$^{87}$Rb$^{133}$Cs molecules have been formed via magnetoassociation in both
Innsbruck~\cite{K.Pilch_pra_2009, T.Takekoshi_pra_2012} and
Durham~\cite{H.Cho_pra_2013, M.P.Koppinger_pra_2014}. The Innsbruck group
\cite{M.Debatin_pccp_2011} subsequently performed detailed one- and two-photon
molecular spectroscopy and, very recently, reported the transfer of molecules
to the rovibrational and hyperfine ground state by
STIRAP~\cite{Takekoshi_arxiv_2014}.

In this letter, we demonstrate STIRAP transfer of $^{87}$Rb$^{133}$Cs molecules
from a bound state near dissociation to the rovibrational ground state,
producing a sample of over 1000~ground-state molecules. In the process we make
detailed measurements of the binding energy and the splitting between the
$J''=0$ and 2 rotational levels of the vibrational ground state ($v''=0$).
Stark spectroscopy of both the excited and ground states is presented, leading
to a precise measurement of the ground-state permanent electric dipole moment.
We demonstrate a space-fixed dipole moment which is substantially larger than
in previous work.

\begin{figure}
	\centering
\includegraphics[width=1.0\linewidth,trim=0.5cm 1cm 0.5cm
0.6cm]{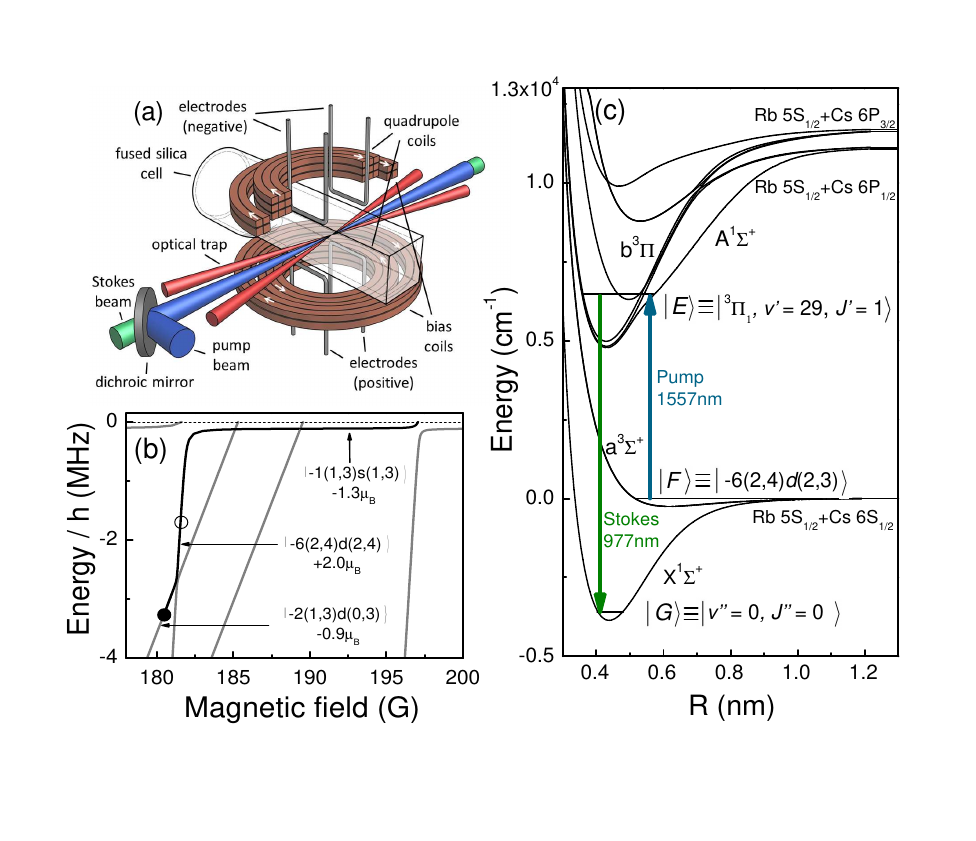}

\caption{\label{figOverview} (Color online) Experimental overview. (a):
Simplified diagram of key elements of the apparatus. (b): Near-threshold bound
states for $^{87}$Rb$^{133}$Cs and the magnetoassociation path (solid black line).
Stern-Gerlach separation is carried out at 180.487(4)~G (closed circle) and
STIRAP transfer is carried out at 181.624(1)~G (open circle). An avoided
crossing at~$\sim$181.3~G allows transfer between these two states. (c):
Potential energy curves for RbCs, indicating the transitions used for STIRAP.}
\end{figure}

Details of the apparatus have been described in our studies of dual-species
condensates \cite{D.J.Mccarron_pra_2011, H.Cho_epjd_2011} and Feshbach
spectroscopy \cite{H.Cho_pra_2013, M.P.Koppinger_pra_2014}. In this work we use
a nearly-degenerate sample of $\sim2.5\times10^5$ $^{87}$Rb atoms in the
$\ket{f=1,m_f=1}$ state and $\sim2.0\times10^5$ $^{133}$Cs atoms in the
$\ket{3,3}$ state, confined in the levitated dipole trap illustrated in
Fig.~\ref{figOverview}(a) at a temperature of ${\sim\unit{300}{\nano\kelvin}}$.
The near-threshold bound states of $^{87}$Rb$^{133}$Cs relevant to our
magnetoassociation sequence are shown in Fig.~\ref{figOverview}(b). As in
\cite{T.Takekoshi_pra_2012}, these states are labeled as
$\ket{n(f_{\text{Rb}},f_{\text{Cs}})L(m_{f_{\text{Rb}}},m_{f_{\text{Cs}}})}$,
where $n$ is the vibrational label for the particular hyperfine
$(f_{\textrm{Rb}},f_{\textrm{Cs}})$ manifold, counting down from the
least-bound state which has ${n=-1}$, and $L$ is the quantum number for
rotation of the two atoms about their center of mass. All states have
$M_{\textrm{tot}}=4$, where ${M_{\textrm{tot}}=M_{F}+M_{L}}$ and
${M_{F}=m_{f_{\textrm{Rb}}}+m_{f_{\textrm{Cs}}}}$.

To create weakly-bound molecules we sweep the magnetic field across a Feshbach
resonance at $\unit{197.10(3)}{G}$ to produce weakly-bound molecules in the
$\ket{-1(1,3)s(1,3)}$ state. These molecules are then transferred to the
$\ket{-2(1,3)d(0,3)}$ state at $\unit{180.487(4)}{G}$ following the path shown
in Fig.~\ref{figOverview}(b) and separated from the remaining atoms using the
Stern-Gerlach effect~\cite{M.P.Koppinger_pra_2014}. To detect the molecules, we
ramp back to a field above the $\unit{197.10(3)}{G}$ resonance to dissociate
the molecules to atoms, which are then detected by absorption imaging. We
typically create trapped samples of $\sim2500$ molecules in the
$\ket{-2(1,3)d(0,3)}$ state with the same temperature as the original atomic
sample and a lifetime of $\unit{200}{\ms}$.

\begin{figure}
	\centering
\includegraphics[width=1.0\linewidth]{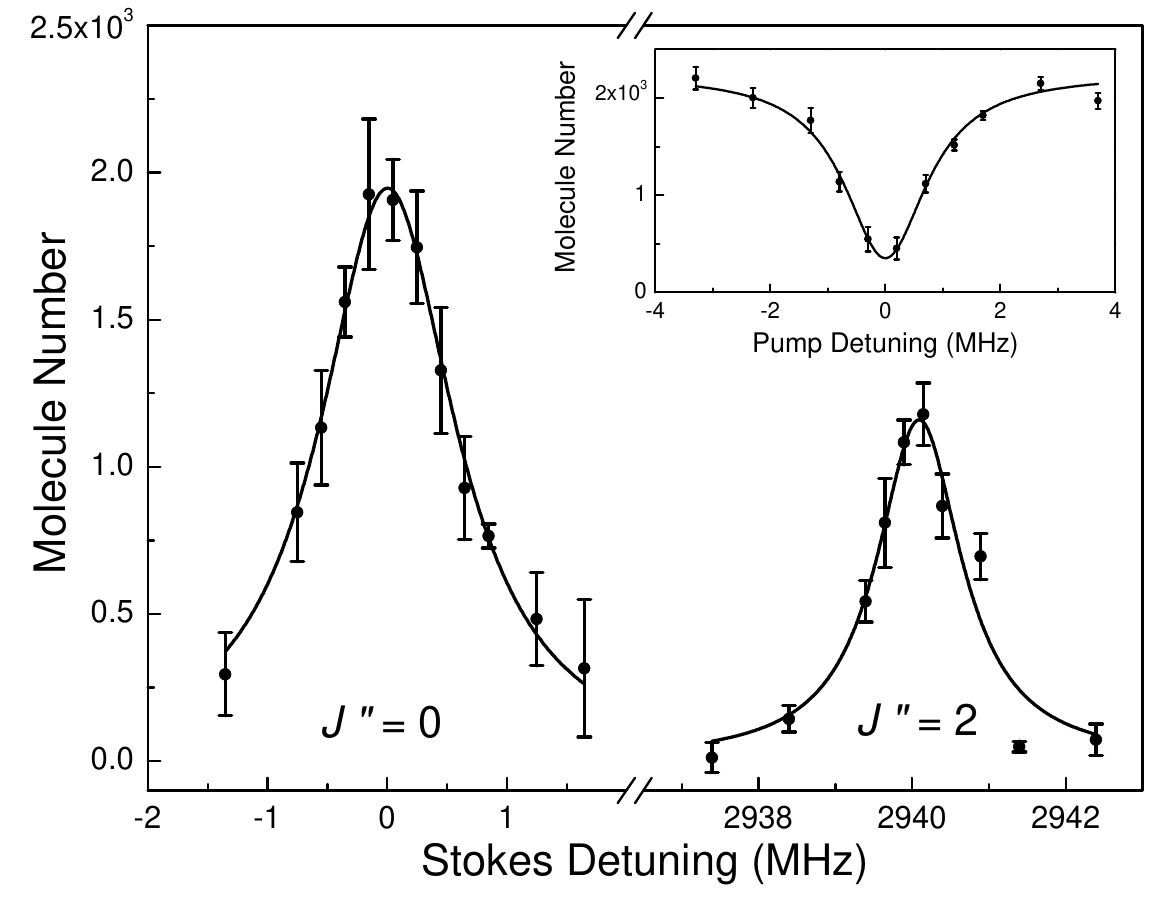}
\caption{Two-photon spectroscopy of the RbCs vibrational ground state. The
molecules remain in the initial near-dissociation state when the Stokes
light is on resonance with the $J''=0$ and $J''=2$ rotational states. The
solid lines illustrate Lorentzian fits used to determine the
resonance positions. Inset: One-photon loss spectrum for the
$\ket{^3\Pi_1,v'=29, J'=1}$ state. The pump laser is held on resonance
with this transition during the two-photon spectroscopy.}
	\label{FigDarkStateSpectroscopy}
\end{figure}

We transfer the molecules from the weakly-bound state~$\ket{F}$ to the ground
state~$\ket{G}$ by coupling them via a level~$\ket{E}$ of the coupled
$A^1\Sigma^{+}+b^3\Pi$ manifold. This requires two lasers, labeled the pump
($\lambda = \unit{1557}{\nm}$) and Stokes ($\lambda = \unit{977}{\nm}$) lasers
in Fig.~\ref{figOverview}(c). These are referenced to an optical
cavity~\cite{Supp}. Up to $\unit{16}{\mW}$ of light at each wavelength can be
focused to a $\unit{\sim 35}{\um}$ waist at the molecule sample. High transfer
efficiency requires a high value of $\Omega^2/\gamma$ for both transitions,
where~$\Omega$ is the Rabi frequency and~$\gamma$ is the natural linewidth.
Debatin \emph{et~al.} have identified several states suitable for
STIRAP~\cite{M.Debatin_pccp_2011}. To locate the states, we pulse
$\unit{20}{\micro\watt}$ of pump light, polarized parallel to the magnetic
field, on the molecules in the~$\ket{F}$ state for $\unit{750}{\us}$. We
observe the molecule loss as a function of optical frequency, as shown in the
inset to Fig.~\ref{FigDarkStateSpectroscopy}, and locate the center with a
Lorentzian fit. In total we have detected seven electronically-excited states
and their numerous hyperfine sublevels spanning a $\unit{1.8}{\tera\hertz}$
range. We focus on the transition to the lowest hyperfine state of the
odd-parity ($e$) component of the $\ket{^3\Pi_1,v'=29, J'=1}$ level, which has
well-separated hyperfine states and high
$\Omega_{\textrm{P}}^2/\gamma$~\cite{M.Debatin_Thesis_2013}.

To determine the pump Rabi frequency $\Omega_{\textrm{P}}$, we set the pump
laser on resonance, vary the pulse time $t$, and fit the fraction of remaining
molecules $N/N_{0}$ to $N/N_{0}=\exp\left({- \Omega_{\textrm{P}}^2
t}/{\gamma}\right)$. We observe stronger coupling to the $\ket{^3\Pi_1,v'=29,
J'=1}$ state from $\ket{-6(2,4)d(2,4)}$ than from $\ket{-2(1,3)d(0,3)}$.
Unfortunately, the former has a positive magnetic moment and cannot be
magnetically levitated in our current setup. We therefore increase the depth of
the optical trap to $\unit{12.7}{\micro\kelvin}$, ramp off the magnetic field
gradient to transfer to an all-optical trap and adjust the bias field to
transfer the molecules into $\ket{F}\equiv\ket{-6(2,4)d(2,4)}$ at
$\unit{181.624(1)}{G}$ (open circle, Fig.~\ref{figOverview}(b)). In this trap
we observe a pump Rabi frequency of $\unit{2\pi\times0.18(1)}{\MHz}$ at maximum
power for the transition from $\ket{F}$ to $\ket{E}\equiv\ket{^3\Pi_1,v'=29,
J'=1}$. An unlevitated trap is advantageous for STIRAP as it removes the
variable Zeeman shift across the cloud. However, the transfer to the deeper
trap heats the molecular cloud to $\unit{1.5(2)}{\micro\kelvin}$ and we observe
a shorter lifetime of $\unit{23(2)}{\ms}$ in state $\ket{F}$.

We find the Stokes transition by setting the pump laser on resonance with an
increased power of $\unit{40}{\micro\watt}$, simultaneously pulsing on
$\unit{16}{\mW}$ of Stokes light polarized perpendicular to the magnetic field
and scanning the optical frequency. At two-photon resonance, a dark state forms
and the state $\ket{F}$ is not excited to the lossy $\ket{E}$ state. This is
observed as an increase in the $\ket{F}$ state population if
$\Omega_{\textrm{S}}\gg\Omega_{\textrm{P}}$, as seen in
Fig.~\ref{FigDarkStateSpectroscopy}. We observe both the $J''=0$ and $J''=2$
levels of the electronic and vibrational ground state, separated by
$h~\times~\unit{2940.09(6)}{\MHz}$. To our knowledge this is the most accurate
direct measurement of this splitting. Neglecting centrifugal distortion, this
implies a rotational constant
$B_0=\unit{0.0163452(3)}{\reciprocal{\centi\metre}}$, which is consistent with
the theoretical prediction of
$\unit{0.0163(4)}{\reciprocal{\centi\metre}}$~\cite{Yang_JPhysChem_2012}. We
measure absolute frequencies of $\unit{192572.09(2)}{\GHz}$ and
$\unit{306830.49(2)}{\GHz}$ for the pump and Stokes transitions respectively.
This implies a zero-field binding energy of
${hc~\times~\unit{3811.576(1)}{\reciprocal{\centi\metre}}}$ for the $J''=0$
state, relative to the degeneracy-weighted hyperfine centers. This is
consistent with the latest theoretical values~\cite{Yang_JPhysChem_2012} and
experimental measurements \cite{M.Debatin_pccp_2011}.

\begin{figure}
\begin{center}
\includegraphics[width=1.0\linewidth]{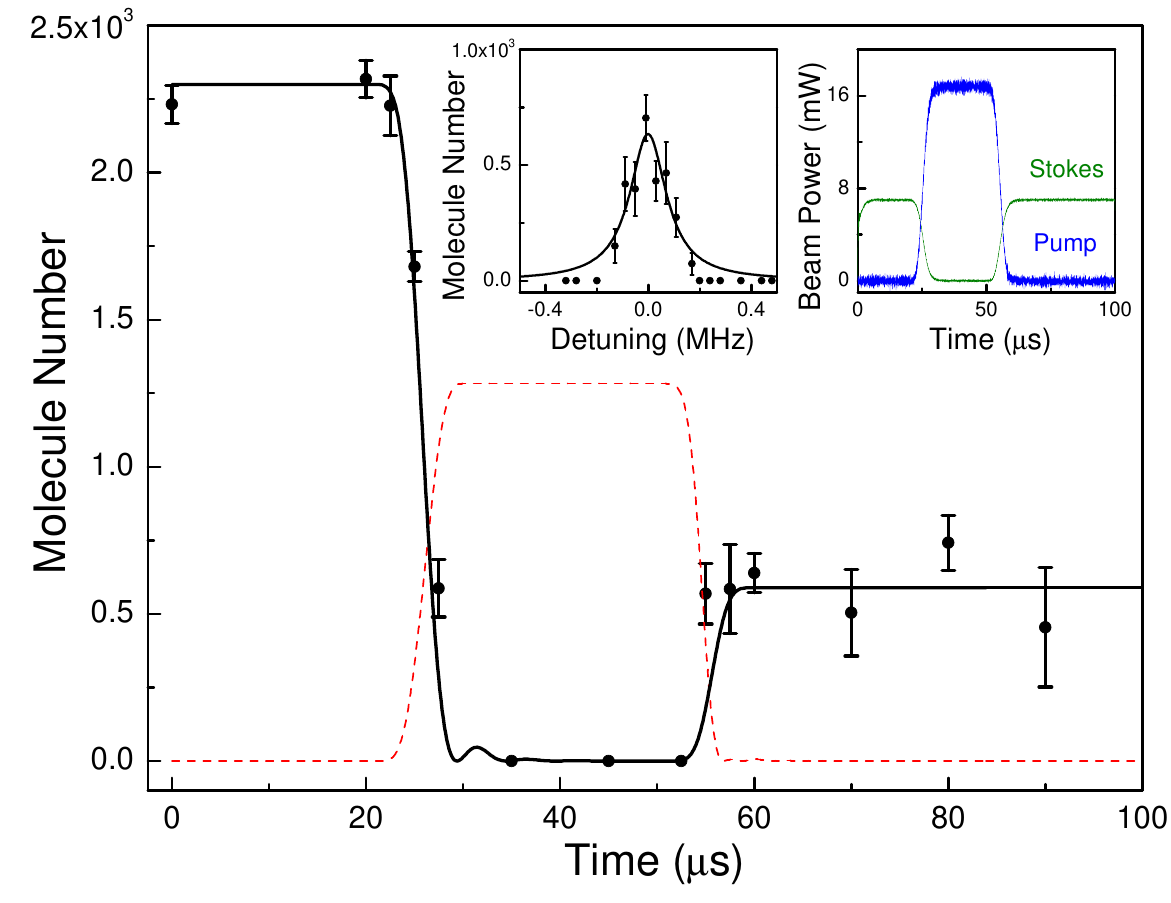}
\caption{(Color online) The number of molecules remaining in the Feshbach state
$|F\rangle$ when both lasers are switched off during the STIRAP sequence. The
black solid and red dashed lines show the Feshbach and ground-state populations
obtained from the Lindblad model described in the text. Left inset: The final
population of Feshbach molecules as a function of Stokes detuning. Right inset:
The pump and Stokes beam powers during the STIRAP pulse sequence.}
	\label{FigSTIRAP}
\end{center}
\end{figure}

Transfer to the ground state via STIRAP relies on a dark state $\ket{D}$ that
is an eigenstate of the system on two-photon resonance. This is composed of a
superposition of the $\ket{F}$ and $\ket{G}$ states,
${\ket{D}=\cos\theta\ket{F}+\sin\theta\ket{G},}$ where the mixing angle
$\theta$ is defined by ${\tan\theta=\Omega_{\textrm{P}}/\Omega_{\textrm{S}}}$.
Transfer from state $\ket{F}$ to $\ket{G}$ (and back) is then achieved by an
adiabatic change in the mixing angle, using the pulse sequence shown in the
right inset to Fig.~\ref{FigSTIRAP}. The Stokes beam is initially turned on to
$\unit{7}{\mW}$ for $\unit{20}{\us}$. With $\Omega_{\textrm{S}}\neq0$ and
$\Omega_{\textrm{P}}=0$, $\ket{D}$ is equivalent to the starting state
$\ket{F}$. The Stokes beam is then ramped down in $\unit{10}{\us}$ while the
pump beam is ramped up to $\unit{16}{\mW}$. This adiabatically transfers the
population to the ground state $\ket{G}$. We cannot detect the ground state
directly, so after a $\unit{20}{\us}$ hold we reverse the process to transfer
back to the initial state, allowing measurement of the square of the one-way
efficiency. The maximum efficiency is achieved with both lasers on resonance,
as shown in the left inset to Fig.~\ref{FigSTIRAP}. We map out the transfer
by truncating the pulse sequence and recording the molecules remaining in the
state $\ket{F}$, as shown in Fig.~\ref{FigSTIRAP} for the on-resonance case.

The polarizations of the pump and Stokes beams drive $\Delta M_{\text{tot}}=0$
and $\pm1$ transitions, respectively. The state $\ket{F}$ has
$M_{\text{tot}}=4$, so that we can reach states with ${M_{\text{tot}}=3}$ or
${M_{\text{tot}}=5}$. Takekoshi~{\em et al.}~\cite{Takekoshi_arxiv_2014} have
shown that the transition to the $M_{\text{tot}}=5$ hyperfine state has the
strongest coupling from $\ket{E}$, with very little population transferred into
the $M_{\text{tot}}=3$ states. Furthermore, theory shows that the $I''=5,
M_{\text{tot}}=5$ state, where $I''$ is the total molecular nuclear spin, is
the lowest hyperfine state at magnetic fields above about
90~G~\cite{Aldegunde_PRA_2008}. We observe only one state (left inset to
Fig.~\ref{FigSTIRAP}) which we therefore conclude is the $M_{\text{tot}}=5$
absolute ground state.


We model the transfer by numerically integrating the Lindblad master
equation~\cite{Supp}. The maximum pump Rabi frequency of
$\unit{2\pi\times0.18(1)}{\MHz}$ is taken from one-photon measurements (see
above) and both detunings are set to zero. The Stokes Rabi frequency is fitted
as a free parameter and thus estimated as $\unit{2\pi\times0.21(1)}{\MHz}$. The
results indicate a one-way transfer efficiency of~$50\%$ and the model shows
that we produce ground-state $\sim1250$ molecules (red dashed line in Fig.~\ref{FigSTIRAP}).
The efficiency is currently limited by the Rabi frequency achieved for the pump
transition. We note that a slower transfer does not increase the efficiency,
because of increased laser frequency noise over longer timescales.

\begin{figure}
\centering
\includegraphics[width=1.0\linewidth]{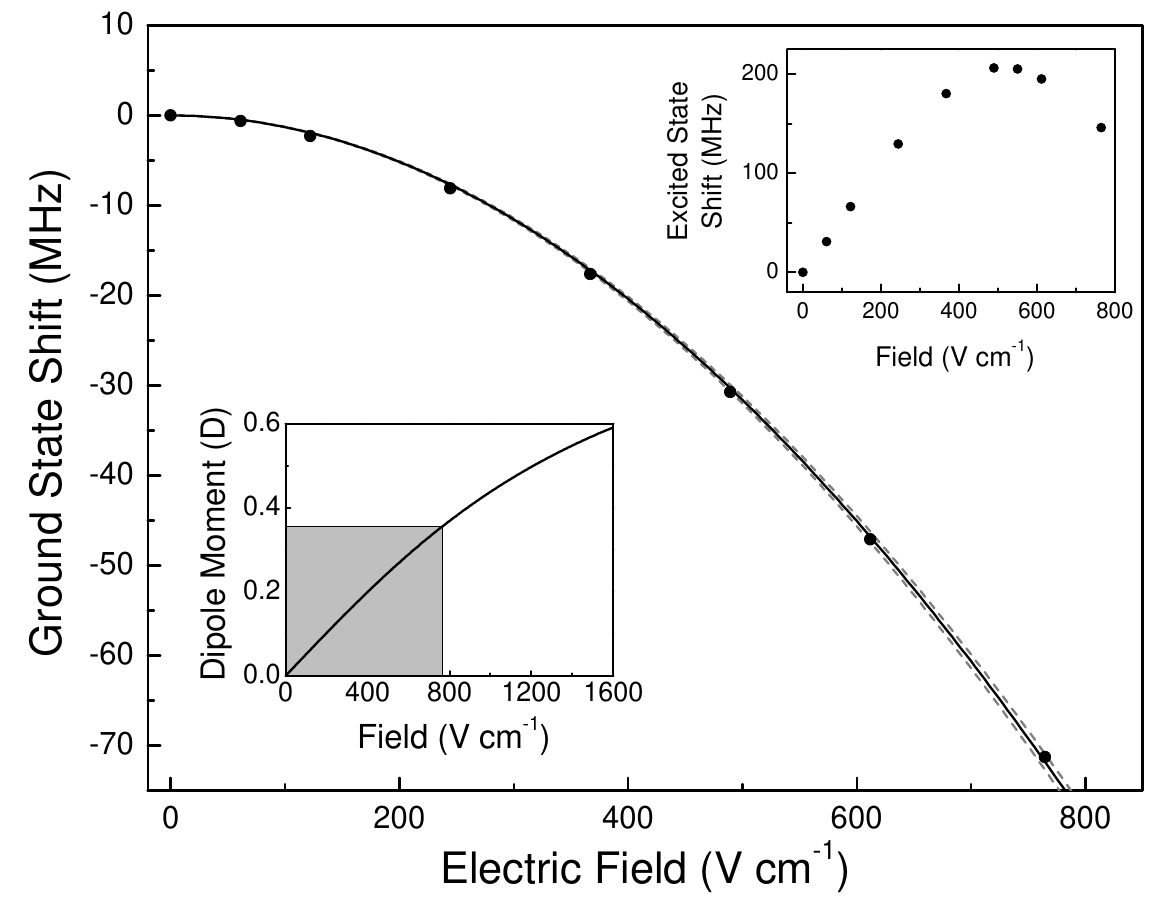}
\caption{Stark shift of the rovibrational ground state. The solid black line
shows the curve fitted to our results from which we extract a permanent
electric dipole moment in the molecular frame of $\unit{1.225(3)(8)}{D}$. The
dotted gray lines indicate the upper and lower bounds due to the systematic
error in the electric field calculation. Upper inset: Stark shift of the
$\ket{^3\Pi_1,v'=29, J'=1}$ excited state used for Stark spectroscopy and
STIRAP. The behavior is initially linear with a gradient of approximately
500~kHz/(V~cm$^{-1}$) up to a field of $\sim 400$~V~cm$^{-1}$. Lower inset:
Ground-state electric dipole moment in the laboratory frame as a function of
electric field. The gray region indicates the range of electric dipole moments
currently accessible in the experiment.}
	\label{FigGroundStark}
\end{figure}

The permanent electric dipole moment of a polar molecule is the key quantity of
interest for many applications. Without an externally applied electric field,
the averaged electric dipole moment in the laboratory frame is zero. Turning on
an electric field couples states of opposite parity and hence polarizes the
molecules in the direction of the field. In the experiment, we apply the
necessary electric field with an array of four electrodes positioned outside
the fused silica cell (shown in Fig.~\ref{figOverview}(a))~\cite{Supp}. We
first measure the DC Stark shift of the pump transition as a function of the
applied electric field. The result is shown in the upper inset in
Fig.~\ref{FigGroundStark}; the initially linear response is interrupted by an
apparent avoided crossing with higher-lying hyperfine states. By following the
avoided crossing, we can then measure the relative shift between the $\ket{E}$
and $\ket{G}$ states (the Stokes shift). As the electric dipole moment of the
state $\ket{F}$ is negligible due to the large interatomic separation, the
difference between the pump and Stokes shifts yields the DC Stark shift of the
rovibrational ground state (shown in Fig.~\ref{FigGroundStark}).

We fit the Stark shift by calculating the matrix solution of the rigid-rotor
Stark Hamiltonian in the laboratory frame~\cite{Supp}. We find a permanent
electric dipole moment in the rovibrational ground state of 1.225(3)(8)~D. The
first uncertainty is statistical and the second is systematic, arising from the
uncertainty in the electric field~\cite{Supp}. Takekoshi~{\em et al.}\ recently
reported a measured value of 1.17(2)(4)~D~\cite{Takekoshi_arxiv_2014}, which
agrees with our measurement within their uncertainty. The lower inset in
Fig.~\ref{FigGroundStark} shows the fitted DC Stark shift converted into the
equivalent electric dipole moment in the laboratory frame, and the gray region
indicates the dipole moment range currently accessible in the experiment. The
maximum laboratory-frame dipole moment we can access is 0.355(2)(4)~D at an
electric field of 765~V~cm$^{-1}$.

Our results complement those reported recently by
Takekoshi~\emph{et~al.}~\cite{Takekoshi_arxiv_2014}. The two experiments
produce a similar number of molecules in the rovibrational ground state, even
though Takekoshi~\emph{et~al.} demonstrate higher STIRAP efficiencies
of~$90\%$. A key difference between the two experiments lies in the trap
geometry; our work uses a simple 3D optical trap, whereas
ref.~\cite{Takekoshi_arxiv_2014} uses a lattice of 2D pancake-shaped traps. A
key feature of our work is that we can apply a larger electric field. This
allows measurement of the ground-state dipole moment with smaller uncertainties
and the realization of larger laboratory-frame electric dipole moments than
in~\cite{Takekoshi_arxiv_2014}.


In conclusion, we have presented high-precision spectroscopy of the ground
state of $^{87}$Rb$^{133}$Cs molecules and demonstrated STIRAP transfer to
create a sample of over 1000~molecules in the rovibrational ground state. The
binding energy of this state is $hc~\times~3811.576(1)$~cm$^{-1}$ and the
splitting between the $J''=0$ and 2 rotational levels of the vibrational ground
state is $h~\times~\unit{2940.09(6)}{\MHz}$. We have used DC Stark spectroscopy
to make a precise measurement of the ground-state permanent electric dipole
moment as 1.225(3)(8)~D, and demonstrate that laboratory-frame dipole moments
up to 0.355(2)(4)~D are accessible in our experiment. We believe that this is
the largest dipole moment in the laboratory frame accessible in any ultracold
molecule experiment to date. For comparison, in $^{87}$Rb$^{133}$Cs,
Takekoshi~\emph{et~al.} access laboratory-frame dipole moments of
$\sim0.03$~D~\cite{Takekoshi_arxiv_2014}, whilst in KRb, Ni~\emph{et~al.}
report values up to 0.22~D~\cite{K.-K.Ni_Nature_2010}. This brings the
possibility of observing strong dipolar interactions in a stable ultracold
molecular gas within reach.

\section{Acknowledgements}
We acknowledge T. Ogden for his help in developing the STIRAP simulation,
D.L. Jenkin, D.J. McCarron and H.W. Cho for their work on the early stages
of the project and useful discussions with H.-C. N\"agerl and members of
his group. This work was supported by the UK EPSRC and by EOARD Grant
FA8655-10-1-3033. CLB is supported by a Doctoral Fellowship from Durham
University. The data presented in this paper are available on request.

\bibliography{References}

\begin{thebibliography}{39}
\bibitem [{Rea()}]{Reactions}%
  \BibitemOpen
  \href@noop {} {\ }\bibinfo {note} {The trimer formation reactions referred to
  here are the two-body reactions
  $2\textrm{RbCs}\rightarrow\textrm{Rb}_2\textrm{Cs}+\textrm{Cs}$ or
  $\textrm{Cs}_2\textrm{Rb}+\textrm{Rb}$}\BibitemShut {NoStop}%
\end{thebibliography}
\end{document}


\section{Supplementary Material}

\subsection{STIRAP laser setup}
The two lasers used for STIRAP are locked to a single cylindrical
$\unit{10}{\centi\metre}$ high-finesse plane-concave optical cavity, built from
ultra-low-expansion glass (ULE) by ATFilms, housed in a vacuum chamber from
Stable Laser Systems and actively stabilised to the zero-expansion temperature
of the ULE at $\unit{35}{\degree\Celsius}$. The properties of the reference
cavity are summarised in Table~\ref{TabCavitySummary}. Light entering the cavity
first passes through a fibre-coupled electro-optic modulator (EOM), adding
tunable sidebands at up to $\unit{10}{\GHz}$ which can span the
$\unit{1.5}{\GHz}$ free spectral range (FSR) of the cavity. Additional sidebands
are added at $\sim 10$\,MHz in order to implement a standard Pound-Drever-Hall
lock to the cavity. By locking to a sideband instead of a fundamental mode, we
can tune the laser frequency continuously between the cavity modes. We measure
the FSR of the reference cavity to $\unit{\pm1}{\KHz}$ by locking to a
TEM$_{00}$ mode and scanning the driving frequency of the fibre EOM across the
adjacent TEM$_{00}$ mode, monitoring the change in the cavity transmission in
the process. Such precise knowledge of the FSR is important for the measurements
of the rotational splitting of the ground vibrational state of the $^{87}$RbCs
molecule. The absolute optical frequencies of both lasers are measured with a commercial
wavemeter, calibrated with the
$5\textrm{P}_{3/2}\leftrightarrow4\textrm{D}_{5/2}$ transition in Rb at
$\unit{196}{\tera\hertz}$.

\begin{table}[h]
	\centering
	\begin{tabular}{l|rr}
		Test wavelength (nm) & 977   & 1557 \\
		\hline
		Zero-expansion ($\unit{}{\degree\Celsius}$) & {35} & $-$\\
		Length($\unit{}{\mm}$) & 100.13958(7) & 100.15369(7) \\
		FSR ($\unit{}{\MHz}$) & 1496.873(1) & 1496.662(1) \\
		Finesse & $1.37(6)\times10^4$ & $1.19(6)\times10^4$ \\
		Mode linewidth ($\unit{}{\KHz}$) & 109(5) & 126(5) \\
	\end{tabular}%
\caption{Properties of the optical cavity used to narrow the linewidths of the
STIRAP lasers.}
	\label{TabCavitySummary}
\end{table}

Fixed frequency ($\unit{80}{\MHz}$) acousto-optic modulators provide continuous
adjustment of the power of each of the STIRAP beams. The beams are coupled
through $\unit{8}{\metre}$ fibres to the main experiment, collimated and
overlapped on a dichroic mirror. The pump and Stokes beams are then focussed by
a single achromatic lens to overlapped $\unit{37(1)}{\um}$ and
$\unit{35.3(6)}{\um}$ spots respectively at the molecular sample.

\subsection{STIRAP transfer modelling}
\begin{figure}[b]
\begin{center}
	\includegraphics[width=0.8\linewidth]{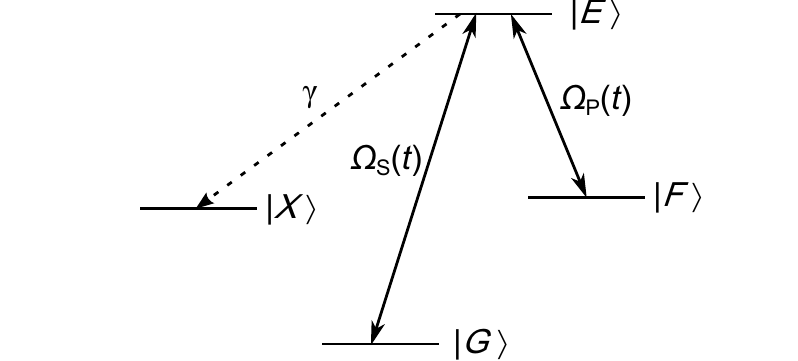}
\caption{Level scheme used in the ground state transfer model. Solid lines are
driven transitions, dashed lines are radiative decays.}
	\label{FigModelDiagram}
\end{center}
\end{figure}
We model the transfer using a four-level scheme as shown in
Fig.~\ref{FigModelDiagram}. The Feshbach state $\ket{F}$, excited state
$\ket{E}$ and ground state $\ket{G}$ are coupled by two Rabi frequencies
$\Omega_{\textrm{P}}(t)$ and $\Omega_{\textrm{S}}(t)$ describing the pump and
Stokes fields, respectively. Loss of population from the system due to the
non-adiabaticity of the transfer is modelled as a decay of $\ket{E}$ to a dump
level $\ket{X}$ at the rate determined by the natural linewidth of the excited
state, $\gamma$. We are currently unable to measure the out-of-loop laser noise
directly and therefore do not include it in the model. We build this system
using the QuTiP module in Python~\cite{Johansson_CPC_2012}, and numerically
integrate the Lindblad master equation. $\Omega_{\textrm{P}}(t)$ and
$\Omega_{\textrm{S}}(t)$ are calculated from an analytic approximation of the
measured response of the beam modulators and the maximum Rabi frequencies.
$\Omega_{\textrm{P,max}}$ is measured as explained in the main text, while
$\Omega_{\textrm{S,max}}$ is fitted as a free parameter. The round-trip transfer
is taken as the simulated population of the $\ket{F}$ state at the end of the
ramp sequence.

We observe that the proximity of the $\lambda=\unit{1550}{\nm}$ trapping lasers
to the pump transition at $\lambda=\unit{1557}{\nm}$ produces a large AC Stark
shift. We also note there is considerable broadening of the pump transitions to
greater than $\unit{1}{\MHz}$, which may be attributable to the variable Stark
shift across the cloud. In our models and calculations we therefore use the
$\unit{130}{\KHz}$ natural linewidth measured by
Takekoshi~\emph{et~al.}~\cite{Takekoshi_arxiv_2014}.

\subsection{Electric field creation and calculation}

The electric field is generated by a set of four electrodes running parallel to
the outside of the fused silica cell. The electrodes are 1.5~mm in diameter,
separated by 29.0(2)~mm vertically, 24.8(2)~mm horizontally and are 22.0(5)~mm
long. They are bent at either end at a radius of 2(1)~mm to allow electrical
connections to be made without interfering with the already limited optical
access to the cell. The electrodes are currently connected in pairs so that the
two electrodes above the cell are negatively charged, and the two below the cell
are positively charged. We do not currently have any experimental method of
calibrating our electric field. We therefore calculate the electric field at the
center of the cell by the method of finite element analysis to solve the Poisson
equation. A 3D mesh is generated from a CAD model of the experimental apparatus
which includes the electrodes, cell walls, earthed magnetic field coils, and
dielectric coil mounts. The mesh creation and subsequent calculation is carried
out in the Autodesk Multiphysics software package. It is found that the presence
of the fused silica cell (external dimensions of
24~mm~$\times$~24~mm~$\times$~80~mm, with 2~mm thick walls) enhances the
electric field by $\sim$2~$\%$. The electric field at the center of the cell
given an applied electric potential of 1~kV between the upper and lower
electrode pairs is 153(1)~V~cm$^{-1}$. The gradient of the electric field at the
center of the cell is zero, so the uniformity is determined by the curvature of
the field as defined by its second order derivative equal to 17(5)~V cm$^{-3}$.
Hence a displacement of the molecular cloud from the center of the cell by 1~mm
leads to a variation in the electric field seen by the molecules of less than
0.1~$\%$.

The high voltage power supplies connected to the electrodes are TTL controlled
so that the field is only applied for the portion of the experiment when the
STIRAP pulses are in operation. It is important for the accurate calculation of
the electric field that we avoid electric polarization of our cell. We find
experimentally that polarization of our cell starts to become detectable at
$\sim$1~kV~cm$^{-1}$. Above this critical field we are able to observe a remnant
electric field with inverse polarity to that previously applied, this causes a
Stark shift of the pump and Stokes transitions even when the electric field is
switched off. We therefore do not make any measurements above 765~V~cm$^{-1}$.
In addition we have measured the Stark shift of the excited state at
245~V~cm$^{-1}$ with the electric field applied in the reverse direction. We
measured a 50~kHz disparity between forward and reverse field Stark shift
measurements, against an overall Stark shift for the transition of 130~MHz.
Remnant polarization can remain in the experiment for days following a high
voltage pulse. To remove this polarization between measurements, we bathe the
cell in UV light. The UV light is generated by a single LED
(Roithner-Lasertechnik LED395-66-60-110), and has a typical output power of
240~mW at 395~nm.

\subsection{Calculation of the fitted ground state DC-Stark shift}

Fitting of the ground-state DC-Stark shift is done by calculating the matrix
elements of the relevant rigid rotor Stark Hamiltonian. These matrix elements
are defined by
\begin{multline}
\langle J~m_{J} \mid H \mid J'~m_{J}' \rangle = B_{0} \cdot J (J+1) \cdot
\delta_{J J' m_{J} m_{J}'} \\ - d_{0}E\sqrt{(2J+1)(2J'+1)} \cdot (-1)^{m_J}
\cdot \\ \left( \begin{array}{c c c}
J & 1 & J' \\
-m_{J} & 0 & m_{J}' \\
\end{array} \right)
\left( \begin{array}{c c c}
J & 1 & J' \\
0 & 0 & 0 \\
\end{array} \right),
\label{eqn:Hamiltonian}
\end{multline}
where $J$ and $m_{J}$ are the rotational quantum numbers and its projection
along the $z$ axis respectively, $B_{0}$ is our experimentally measured
rotational constant, $E$ is the applied electric field, and the two matrices are
Wigner 3-j coefficients. We include rotational levels up to ${J=8}$. This equation
can be understood by considering its two terms separately. The first is the
rotational splitting without the presence of an applied field, the second is the
DC-Stark shift.

\subsection{Systematic error budget for experimentally measured permanent
electric dipole moment in the lab frame}

The major sources of systematic error in the calculated value of the permanent
electric dipole moment can be seen in Table~\ref{table:Uncertainty}. They are
broken down into three areas. The first is due to the uncertainty in the
measurement of the electrode separations, including any uncertainties in the
electrode shape (i.e. the bend radius at either end of each electrode). The next
source of error is caused by the uncertainty in the position of the molecular
cloud with respect to the center of the electrode array. This uncertainty is
estimated to be $\pm1$~mm in any direction. Note that the position of the glass
cell with respect to the electrode array has an error of similar magnitude.
However, calculations of the electric field where we move the glass cell by
$\pm1$~mm show negligible deviation from the central value. The final source of
error is from the electric field calculation itself. Finite element analysis is
an approximate method which relies upon a converging solution. This convergence
has some noise, which is estimated by repeating the calculation with a various
mesh densities. We estimate the uncertainty in this value to be $\pm0.7~\%$ of
the central electric field at a given potential. It is clear that the
uncertainty is dominated predominantly by the accuracy of the measurement of the
electrode position.

\begin{table}[h]
\vspace{0.5cm}
\begin{tabular}{r c}
\hline
~															&				$\alpha_{d_{0}}$~(D)	\\
\hline
Electrode separation					&				$\pm 0.007$				\\
Molecule position							&				$\pm 0.002$				\\
Field calculation							&				$\pm 0.004$				\\
\hline
{\bf Total systematic error}	&				$\pm 0.008$				\\
\hline
\end{tabular}
\caption{\label{table:Uncertainty} Breakdown of the various sources of
systematic uncertainty in the measured value of the permanent electric dipole
moment ($\alpha_{d_{0}}$). }
\end{table}